\documentclass[twocolumn,showpacs,preprintnumbers,amsmath,amssymb, prb]{revtex4}
\usepackage{graphicx}% Include figure files
\usepackage{dcolumn}% Align table columns on decimal point
\usepackage{bm}% bold math
\usepackage{wasysym}
\usepackage{textcomp}
\usepackage{epstopdf}
\usepackage{stmaryrd}
\usepackage{txfonts}
\begin{document}

\preprint{APS/123-QED}

\title{Search and identification of repeated magnetic compensation in Nd$_{1-x}R_x$Al$_2$ ($R$ = Gd, Tb) series}% Force line breaks with \\
\author{P. D. Kulkarni, A. Thamizhavel, S. Ramakrishnan and A. K. Grover}
\affiliation{Department of Condensed Matter Physics and Materials
Science, Tata Institute of Fundamental Research, Homi Bhabha Road,
Colaba, Mumbai 400 005, India.}
\date{\today}% It is always \today, today,

\begin{abstract}
We undertook investigations in polycrystalline $R_x$$R'_{1-x}$Al$_2$ alloys, in particular, explored the stoichiometries, Nd$_{0.8}$Gd$_{0.2}$Al$_2$ and Nd$_{0.8}$Tb$_{0.2}$Al$_2$, to study the magnetic compensation behavior in them in the context of contemporary interest. In these admixed rare-earth (RE) intermetallics, the compensation phenomenon is characterized by a crossover of the $M$ = 0 axis in the low field thermomagnetic response, and a turnaround behavior due to the field-induced reversal in the orientations of the RE moments at high fields. While the characteristic attributes are present in Nd$_{0.8}$Gd$_{0.2}$Al$_2$, a repeated/multiple magnetic compensation behavior stands observed in Nd$_{0.8}$Tb$_{0.2}$Al$_2$. The ac susceptibility, the heat capacity and the transport measurements in the latter compound reveal the onset of two quasi-antiferromagnetic transitions one after the other on cooling down from the high temperature end. The two magnetic transitions involving competition between the contributions from nearly balanced weighted averages of moments of Nd$^{3+}$ and Tb$^{3+}$ ions generate the curious multiple crossings of the $M$ = 0 axis. The results are compared and contrasted with repeated compensation behavior reported earlier in admixed Nd$_{0.8}$Gd$_{0.2}$Rh$_3$B$_2$ alloy.

\end{abstract}

\pacs{71.20.Lp, 75.40.-s, 75.50.Ee}% PACS, the Physics and Astronomy

\keywords{Nd$_{1-x}$R'$_{0.2}$Al$_2$, antiferromagnetism, magnetic compensation, phase transition}

\maketitle
\section {Introduction}
Rare earth ($R$) elements in the 4$f$-Lanthanide series in the periodic table can be combined easily with the nonmagnetic aluminium to form $R$Al$_2$ alloys having the cubic $C15$ Laves phase structure whose basic prototype is the MgCu$_2$ compound\cite{Wallace}. These $R$Al$_2$ alloys have been studied in the last five decades \cite{Purwins} for the detailed information on their magnetic behavior. In the ferromagnetic series of $R$Al$_2$ alloys ($R$ = Pr-Sm, Gd-Er), the indirect exchange coupling between the rare earth moments is governed by the Rudermann-Kittel-Kasuya-Yoshida (RKKY) interaction mediated by the conduction electron spins\cite{Wallace}. Effective moments in the paramagnetic regions for these alloys are in good agreement with the $g_J$($J$($J$+1))$^{1/2}$ values for the free rare earth ions, where $g_J$ is the Lande\textquoteright s $g$-factor and $J$ is the total angular momentum of the $R$-ions. Following Hund's rule, the total ground state eigen value of $J$ can be obtained as $L$ - $S$/$L$ + $S$ for the first/second half elements of 4$f$-series, respectively, and the net magnetic moment is given by $\mu$ = -$g_J\mu_{\rm B}J$, where $\mu_{\rm B}$ is the Bohr magneton. Admixture of two rare earth ions to form $R_{1-x}$$R'_{x}$Al$_2$ alloys generates isostructural and isoelectronic ternary alloys, where the two $R$-ions randomly occupy the sites on the diamond sub-lattice in the $C15$ cubic structure \cite{Wallace}. The magnetic measurements in the admixed solid solutions of rare earth ions in dialuminides were first reported by Williams $et$~$al.$\cite{Williams} in 1962. They reported that for light-light and heavy-heavy combination of rare earth ions, the ferromagnetic ordering prevails by citing the examples Pr$_{1-x}$Nd$_x$Al$_2$, Dy$_{1-x}$Gd$_x$Al$_2$, Er$_x$Gd$_{1-x}$Al$_2$ alloys. In light-heavy combination, e.g., Nd$_{1-x}$Gd$_x$Al$_2$, Pr$_{1-x}$Gd$_x$Al$_2$, Nd$_{1-x}$Tb$_x$Al$_2$, the magnetic moments are coupled antiferromagnetically, which was shown to produce the compensation points in the magnetization versus temperature data in Pr$_{1-x}$Gd$_x$Al$_2$ alloy with $x$ = 0.2 and 0.314. Swift $et$~$al.$\cite{Swift} extended these studies further and explored many light-light, light-heavy and heavy-heavy alloys belonging to $R_{1-x}$$R'_{x}$Al$_2$ series \cite{Wallace}. Grover $et$~$al.$\cite{Grover1} reported the hyperfine field studies on the Sm$_{1-x}$Gd$_x$Al$_2$ alloys, which confirmed microscopically the ferromagnetic interaction between the rare earth spins in the admixed alloys. In more studies in Sm$_{1-x}$Gd$_x$Al$_2$ they also observed\cite{Grover2} the compensation points for the concentrations, $x$ = 0.01, 0.02 and 0.03. Adachi $et$~$al.$\cite{Adachi1, Adachi2, Adachi3, Adachi4} explored several Samarium based intermetallics from the point of view of large spin polarization and near-zero net bulk magnetization and articulated their potential for applications in spintronics/other devices\cite{Adachi1}. 

Recently, Kulkarni $et$~$al.$\cite{Kulkarni} investigated Nd$_{1-x}$Gd$_x$Rh$_3$B$_2$ series of alloys, which have CeCo$_3$B$_2$-type hexagonal structure. The magnetic moments in NdRh$_3$B$_2$ ($T_c$ $\sim$ 15 K) and GdRh$_3$B$_2$ ($T_c$ $\sim$ 90 K) are known to be $\sim$ 2.5 $\mu_{\rm B}$/f.u. and $\sim$ 7.7 $\mu_{\rm B}$/f.u., respectively\cite{Obiraki}. In the admixed series, the ferromagnetic coupling between the spins of Nd$^{3+}$ and Gd$^{3+}$ ions caused their magnetic moments to align antiferromagnetically. It was observed that while the magnetic compensation got easily achieved for 0.20 $\leq$ $x$ $\leq$ 0.25, a novel repeated/multiple magnetic compensation attribute was discovered in the thermomagnetic response for $x$ = 0.25\cite{Kulkarni}. In order to explore the possible occurence of such an interesting magnetic behaviour in $R_{1-x}$$R'_{x}$Al$_2$ alloys with cubic crystal structure, we carried out the investigations in the admixed stoichiometries Nd$_{1-x}$$R_x$Al$_2$ ($R$ = Gd, Ho and Tb). Interestingly, in Nd$_{0.8}$Tb$_{0.2}$Al$_2$, we have witnessed crossing of the $M$ = 0 axis in a multiple manner in the low field thermomagnetic response. These studies fortify the main conclusion of our earlier explorations\cite{Kulkarni} in the Nd$_{1-x}$Gd$_x$Rh$_3$B$_2$. However, the magnetic compensation at higher temperature in the latter series was thought to be linked to the possible quasi-one-dimensional \cite{Obiraki} fluctuation effects along the $c$-axis of hexagonal CoCo$_3$B$_2$ structure, which cannot be the situation in the cubic $C$15 structure for $R$Al$_2$ compounds. We present here the details of the new results of repeated magnetic compensation phenomenon in the admixed Nd$_{0.8}$Tb$_{0.2}$Al$_2$, while making comparison with the iso-structural and iso-electronic Nd$_{0.8}$Gd$_{0.2}$Al$_2$ where only one compensation temperature is observed. We also report prelimenary magnetization data at few values of the applied field in a single crystal sample of Nd$_{0.8}$Tb$_{0.2}$Al$_2$, which confirms the presence of the repeated magnetic compensation identified in the polycrystalline sample. 

\section{Experimental}
Polycrystalline samples of Nd$_{1-x}$$R_x$Al$_2$ ($R$ = Gd and Tb) were prepared by repeatedly melting together the stoichiometric amounts of the constituent elements in an argon arc furnace. Small portions of the final ingots were powdered for the x-ray diffraction studies and the patterns were obtained using Panalytical X'pert powder diffractometer. Figure 1 shows the $\rm{x}$-ray diffraction pattern (using Cu K$\alpha$) in the powdered sample of Nd$_{0.8}$Tb$_{0.2}$Al$_2$ alloy, along with those in the parent alloys, NdAl$_2$ and TbAl$_2$. The x-ray peaks in all the three patterns are normalized with respect to their respective (311) peak intensity. All the x-ray peaks could be indexed to the cubic Laves phase ($C$15) in the three samples. Similar $\rm{x}$-ray patterns are obtained in the powdered specimen of Nd$_{0.8}$Gd$_{0.2}$Al$_2$ and other stoichiometries, on which results are confirmed in this work. The $\rm{x}$-ray lines in the admixed alloys were sharp and at intermediate positions to those corresponding to the parent alloys (cf. inset in Fig. 1). An elemental analysis was also performed using JEOL JSX-3222 system and the targeted stoichiometries were confirmed in the samples. 

The dc magnetization data were recorded using Quantum Design (QD) Inc. superconducting quantum interference device (SQUID) magnetometers (Models MPMS-5 and SQUID-VSM) and a Oxford Instruments vibrating sample magnetometer (VSM). The ac-susceptibility data in the samples were recorded using QD Inc. Model MPMS-5. The heat capacity was measured in a physical property measurement system (QD Inc. Model PPMS). Electrical resistance was recorded using a home made resistivity set up. 

\begin{figure}[h]
\includegraphics[width=0.45\textwidth]{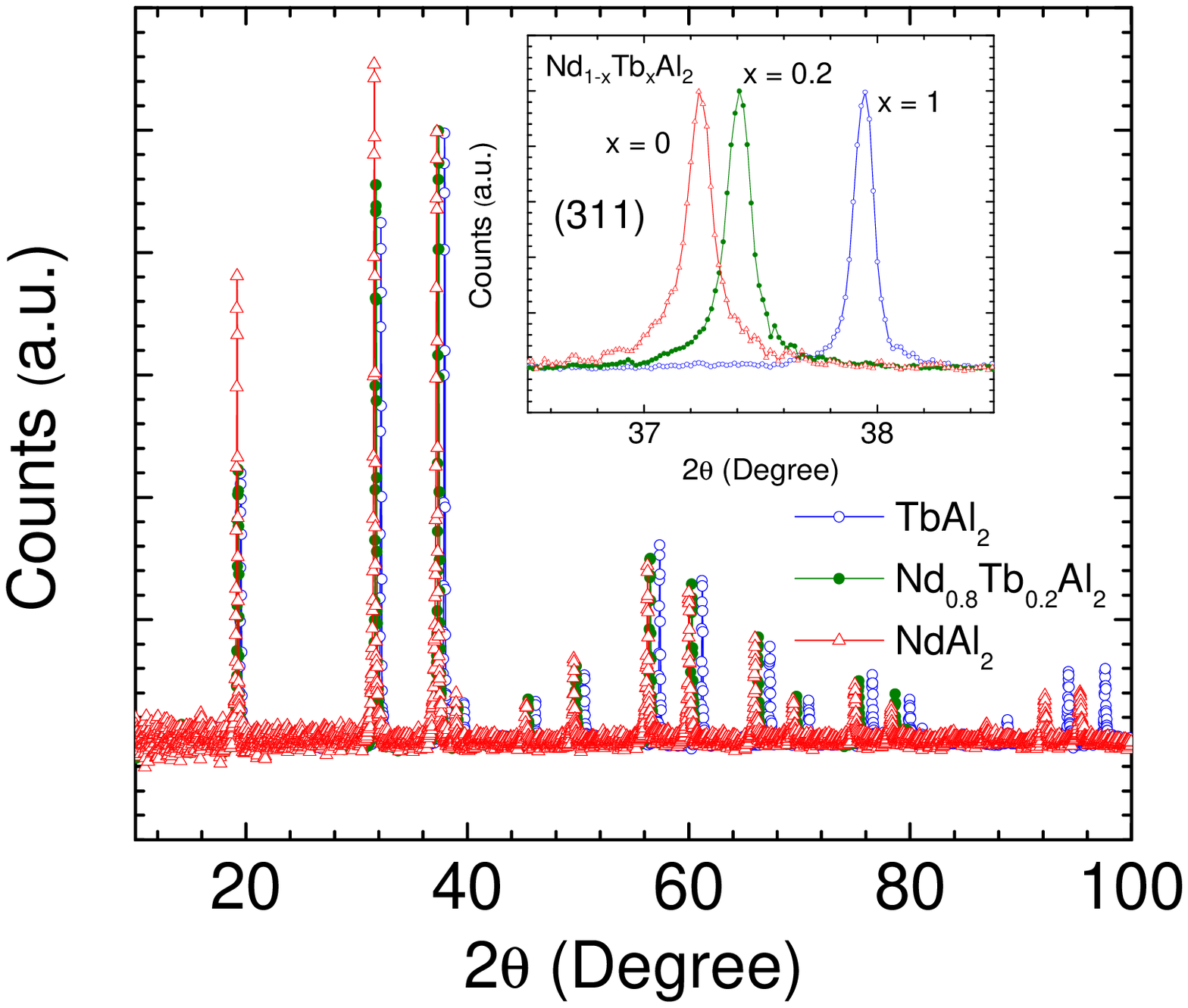}
\caption{\label{Fig1}(Color online) Comparison of powder $\rm{x}$-ray diffraction pattern in Nd$_{0.8}$Tb$_{0.2}$Al$_2$ with those in the samples of NdAl$_2$ and TbAl$_2$. The spectra were obtained using Cu K$\alpha$ 1.5406 $\AA$. An inset panel focuses attention onto the width and location of (311) $\rm{x}$-ray peak in the admixed stoichiometry and the parent compounds in the series Nd$_{1-x}$Tb$_{x}$Al$_2$.}
\end{figure}

\section{Results}
\subsection{Magnetic compensation behavior in dc magnetization data in Nd$_{0.8}$Gd$_{0.2}$Al$_2$ and Nd$_{0.8}$Tb$_{0.2}$Al$_2$}

Figures 2(a) and 2(b) show the temperature variations of field cooled cooldown (FCC) magnetization ($M_{FCC}$($T$)) measured in low field (viz., 100 Oe) and high field (10 kOe) in  Nd$_{0.8}$Gd$_{0.2}$Al$_2$ and Nd$_{0.8}$Tb$_{0.2}$Al$_2$, respectively. The magnetic ordering temperature ($T_c$) in the respective samples have been identified as the onset of very steep rise in magnetization values on cooling down from the paramagnetic end. In a field of 100 Oe, $M_{FCC}$($T$) curves in the two specimen show crossover of $M$ = 0 axis at 74 K and 72 K, respectively, which have been designated as the magnetic compensation temperatures ($T_{comp}$ / $T_{comp1}$). In a much larger field of 10 kOe, the $M_{FCC}$($T$) curves in Figs. 2(a) and 2(b) display the turnaround characteristic at temperatures (identified by $T^*$ / $T_1$$^*$), which are a little higher than the $T_{comp}$ values in the respective panels. The differences in the $T_c$ values of the two admixed stoichiometries can be ascribed to the difference in the $T_c$ values of GdAl$_2$ ($T_c$ $\approx$ 170 K) and TbAl$_2$ ($T_c$ $\approx$ 110 K). The fortuitous closeness in the $T_{comp}$ values of the two specimen can also be rationalized in terms of the differences in $\mu$/f.u. of TbAl$_2$ ($\approx$ 10 $\mu_{\rm B}$/f.u.) and that of GdAl$_2$ ($\approx$ 7.7 $\mu_{\rm B}$/f.u.) and the differences in the temperature dependences of magnetic moments of $L$-$S$ coupled Tb$^{3+}$ and $S$-state ($L$ = 0) Gd$^{3+}$ ions below the respective magnetic ordering temperatures of the admixed compounds. An additional very striking feature in the $M_{FCC}$($T$) curves in 100 Oe/10 kOe in Nd$_{0.8}$Tb$_{0.2}$Al$_2$ is the turnaround behavior in the neighbourhood of 34 K (designated as $T_{f2})$. This indicates the onset of second transformation in the magnetic response of the compound well separated from its $T_{comp1}$ value ($\approx$ 72 K). 

\begin{figure}[h]
\includegraphics[width=0.47\textwidth]{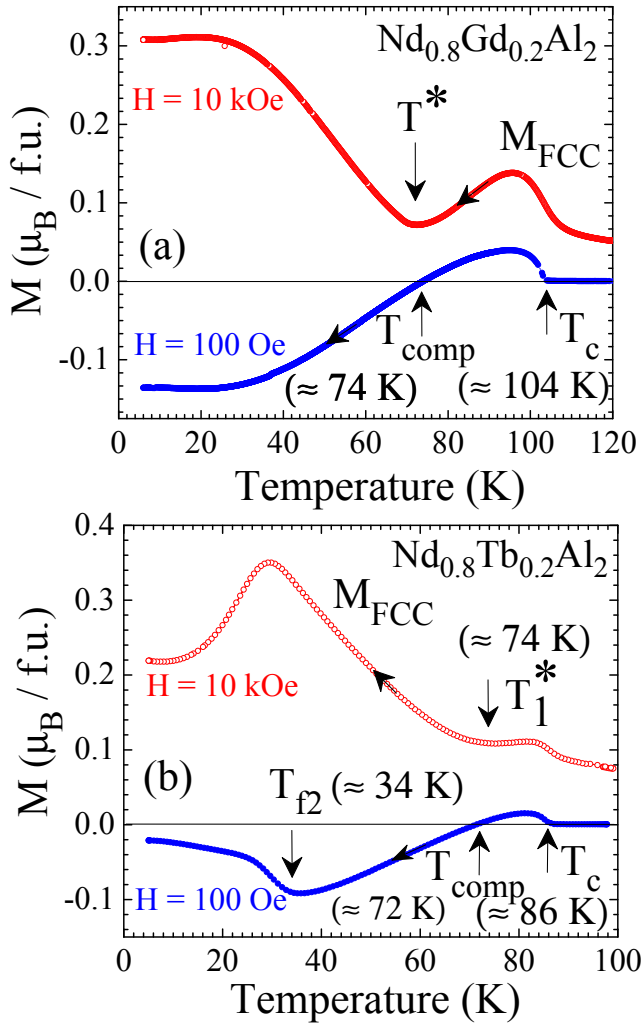}
\caption{\label{fig2}(Color online) Temperature variation of the field cooled magnetization in (a) Nd$_{0.8}$Gd$_{0.2}$Al$_2$ and (b) Nd$_{0.8}$Tb$_{0.2}$Al$_2$ in small field (= 100 Oe) and in high field (10 kOe). $T_c$ value is marked in the low field cool down curve of each of the alloy. The zero crossover of magnetization is observed at temperatures marked as compensation temperatures ($T_{comp}$ / $T_{comp1}$). At higher values of applied magnetic field, a turnaround in the thermomagnetic curve is seen (at $T$* / $T_1$*) due to reversal in the orientations of the antiferromagnetically linked magnetic moments of the dissimilar rare earth ions.}
\end{figure}

To explore the above stated transformation further, we recorded the temperature dependences of the remanent magnetization in Nd$_{0.8}$Gd$_{0.2}$Al$_2$ and Nd$_{0.8}$Tb$_{0.2}$Al$_2$. The sequence of measurements was as follows: Cool a given sample in a field of 10 kOe to 5 K, reduce the field to remanent value of 50 Oe / 100 Oe and then measure the magnetization ($M_{rem}$) in the warm up mode to temperatures above $T_c$; followed by magnetization measurements in the cool down mode ($M_{FCC}$) in the same field to 5 K. Figures 3 (a) and 3 (b) show a comparison of $M_{rem}$($T$) and $M_{FCC}$($T$) in Nd$_{0.8}$Gd$_{0.2}$Al$_2$ and Nd$_{0.8}$Tb$_{0.2}$Al$_2$, respectively. The two sets of curves in each panel appear a mirror image of each other at all temperatures, but for that in the close proximity of $T_c$, and these can be easily understood in terms of the metastable nature of the $M_{FCC}$($T$)/$M_{rem}$($T$) curves at low fields below/above the respective $T_{comp}$ values, keeping in view the large magneto-crystalline anisotropy of Nd$^{3+}$ and Tb$^{3+}$ ions. A noticeble feature, however, is a little enhanced sharpness in the rise, followed by a drop, in the $M_{rem}$($T$) values as $T$ $\rightarrow$ 34 K in Fig. 3(b). This reinforces the notion of an underlying magnetic transition in close proximity of this temperature. More revelations about it emerged from the ac-susceptibility, electrical resistance and specific heat data described below.

\begin{figure}[h]
\includegraphics[width=0.47\textwidth]{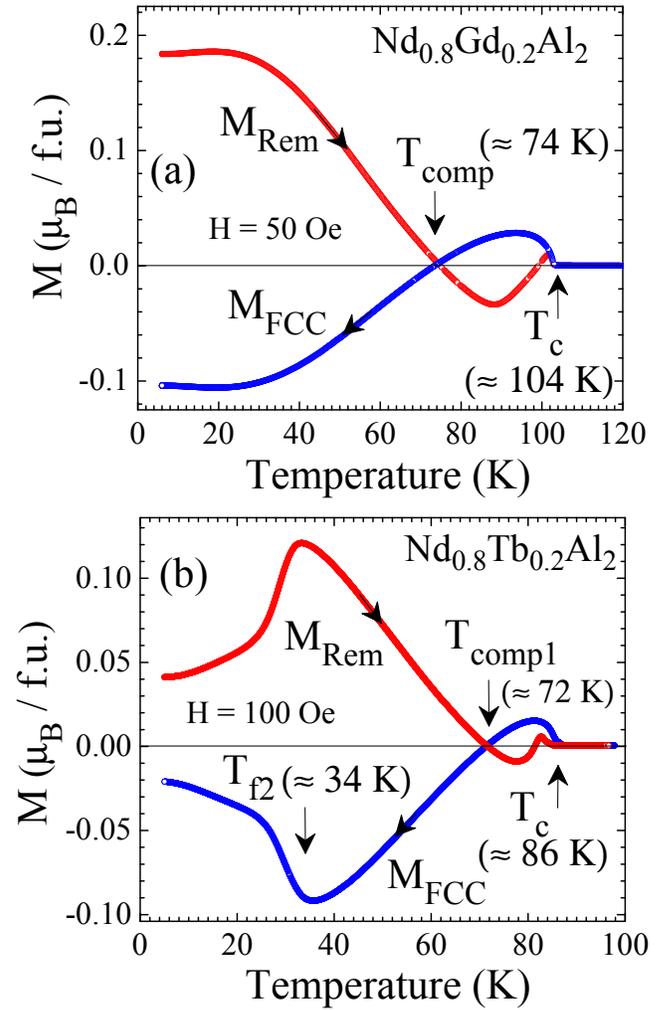}
\caption{\label{fig3}(Color online) The warm up behaviour of the remanent magnetization along with the FCC curve in the same field in (a) Nd$_{0.8}$Gd$_{0.2}$Al$_2$ and (b) Nd$_{0.8}$Tb$_{0.2}$Al$_2$. The remanence at 5 K is obtained by cooling each of the alloys in 10 kOe and reducing the field down to preselected 50 Oe/100 Oe value. The remanence warm up curve intersects with the FCC curve at $T_{comp}$/$T_{comp1}$ and merges with it near $T_c$ in both the alloys.}
\end{figure}

\subsection{AC susceptibility, Electrical Resistance and Specific heat in Nd$_{0.8}$Tb$_{0.2}$Al$_2$}

Figures 4 (a) and 4 (b) summarize the ac-susceptibility data recorded in applied dc fields of $\sim$ 1 Oe, 1 kOe and 10 kOe and the dc electrical resistance ($R$) in earth's field in Nd$_{0.8}$Tb$_{0.2}$Al$_2$, respectively. The two peaks in the in-phase ac susceptibility $\chi$\textquotesingle($T$) curve in nominal zero field ($\sim$ 1 Oe) in Fig. 4(a) reveal the occurence of two magnetic transitions in this sample. The $R$($T$) data in Fig. 4(b) also seem to feebly register the lower temperature transtion near 34 K, in addition to fingerprinting the sharp drop in spin-disorder resistivitiy at $T_c$ ($\approx$ 86 K). It is fruitful to recall here that the two transitions in $\chi$\textquotesingle($T$) and $R$$(T)$ also stand reported\cite{Kulkarni} in the admixed Nd$_{0.75}$Gd$_{0.25}$Rh$_{3}$B$_2$ system, in which the repeated magnetic compensation behavior was recognized, perhaps, for the first time. There are, however, subtle differences in the $\chi$\textquotesingle($T$) and $R$$(T)$ responses in Nd$_{0.8}$Tb$_{0.2}$Al$_2$ and Nd$_{0.75}$Gd$_{0.25}$Rh$_{3}$B$_2$; in the latter system, the fingerprints of change at the higher temperature (i.e., near $T_c$) were very feeble, and a motivated search had to be made for them. 

The $\chi$\textquotesingle($T$) plots in $H$ = 1 kOe and $H$ = 10 kOe in Nd$_{0.8}$Tb$_{0.2}$Al$_2$ in Fig. 4(a) reveal that the two transformations (at $T_c$ and near $T_{f2}$ of 34 K) are robust under the application of dc fields upto 10 kOe.

\begin{figure}[h]
\includegraphics[width=0.45\textwidth]{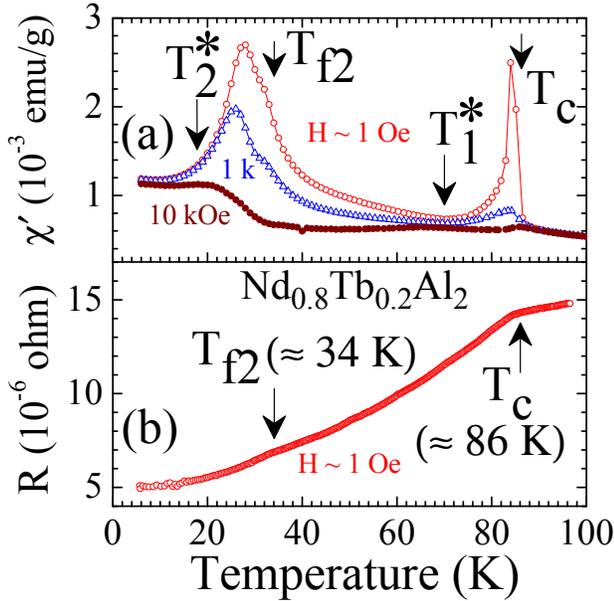}
\caption{\label{fig5}(Color online) (a) AC-susceptibility and (b) electrical resistance ($R$) data as a function of temperature in Nd$_{0.8}$Tb$_{0.2}$Al$_2$ alloy. AC-susceptibility shows the two-peaks below $T_c$ of 86 K and $T_{f2}$ of 34 K. The $R$($T$) in Fig. 4(b) shows changes in the slope of the curve near $T_c$ and $T_{f2}$.}
\end{figure}

In the case of Nd$_{0.75}$Gd$_{0.25}$Rh$_{3}$B$_2$, the in-field specific heat data had confirmed the occurrence of two quasi-antiferromagnetic transitions at higher temperature of $T_{f1}$ (the so called $T_c$ here) and the lower temperature of $T_{f2}$. Figure 5 summarizes the specific heat ($C_p$) data in Nd$_{0.8}$Tb$_{0.2}$Al$_2$ recorded in nominal zero field, 10 kOe and 50 kOe, respectively. The presence of a sharp peak in $C_p$($T$) (in nominal zero field) near $T_c$ can be immediately noted in the main panel of Fig. 5. The temperature region enclosed in the rectangular box in the main panel of Fig. 5 carries in it the fingerprint of transformation initiating near 34 K in the specific heat data. An inset (a) in Fig. 5 shows the effect of application of larger fields of 10 kOe and 50 kOe on the peak in $C_p$($T$) near $T_c$. The characteristic that the peak appears to shift towards lower temperature on application of field signifies\cite{Ramirez} the anti-ferromagnetic nature of the underlying magnetic order setting in at $T_c$. This is analogous to the behavior noted at $T_{f1}$ in Nd$_{0.75}$Gd$_{0.25}$Rh$_{3}$B$_2$. The inset (b) in Fig. 5 shows a comparison of \textquotedblleft excess \textquotedblright\ specific heat, below 40 K, in zero field and in 10 kOe in Nd$_{0.8}$Tb$_{0.2}$Al$_2$. The \textquotedblleft excess \textquotedblright\ specific heat is determined as follows. In a much higher field of 50 kOe, there is no anomalous change in $C_p$($T$) response in the boxed region of the main panel of Fig. 5. We, therefore, plotted the differences in $C_p$($T$) response, $\triangle$$C_p$($T$,$H$), (= $C_p$($T$,$H$) - $C_p$($T$,$H$ = 50 kOe)) in $H$ = 0 Oe and $H$ = 10 kOe in the inset panel (b) of Fig. 5. The peaks in $\triangle$$C_p$($T,H$) are well evident in $H$ = 0 and 10 kOe in this inset panel below $T_{f2}$ of 34 K. The peak height in $H$ = 10 kOe appears to get suppressed vis-a-vis that in $H$ = 0 Oe, and it also shifts somewhat towards the lower temperature side. These characteristics would naively support the quasi-antiferromagnetic nature of the corresponding second transformation getting triggered near $T_{f2}$ as well, as reported earlier in the case of Nd$_{0.75}$Gd$_{0.25}$Rh$_{3}$B$_2$\cite{Kulkarni}. 

\begin{figure}[h]
\includegraphics[width=0.45\textwidth]{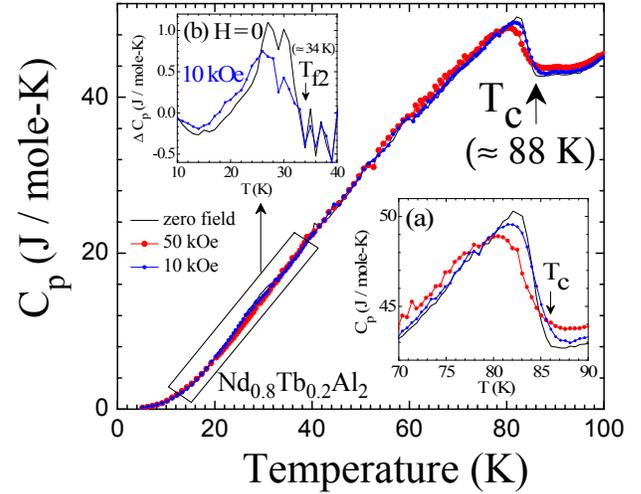}
\caption{\label{fig6}(Color online) Temperature variation of the specific heat ($C_p$) in Nd$_{0.8}$Tb$_{0.2}$Al$_2$ alloy. Fingerprints of two transformations (at $T_c$ and $T_{f2}$) are present in the heat capacity data. The in-field measurements show the shift in the peak positions to the lower temperatures in the inset panels (a) and (b), indicating the two underlying antiferromagnetic transitions in the alloy.}
\end{figure}

\subsection{Repeated magnetic compensation behavior in Nd$_{0.8}$Tb$_{0.2}$Al$_2$} 

The similarities in the several features seen in the magnetic, electrical transport and specific heat data in Nd$_{0.8}$Tb$_{0.2}$Al$_2$ and Nd$_{0.75}$Gd$_{0.25}$Rh$_{3}$B$_2$ motivated us to search for the conspicuous fingerprint of possible occurence of repeated magnetic compensation in the temperature dependent dc magnetization response in the former system, as had been noted earlier in the latter system\cite{Kulkarni}. Figure 6 displays the $M_{FCC}$($T$) curves in $H$ = 225 Oe, 250 Oe and 275 Oe in Nd$_{0.8}$Tb$_{0.2}$Al$_2$ alloy. Topologically, these curves appear similar to $M_{FCC}$($T$) curve in $H$ = 100 Oe in Fig. 2(b). However, one can straight away notice the emergence of the fingerprint of second magnetic compensation, which is triggered by the onset of the lower temperature transformation ($T_{f2}$) below 40 K. Multiple crossings of the $M$ = 0 axis by $M$$_{FCC}$($T$) curve in $H$ = 250 Oe could be viewed as analogous to the multiple crossings of the $M$ = 0 axis in Nd$_{0.75}$Gd$_{0.25}$Rh$_{3}$B$_2$ at low fields ($H$ \textless\ 150 Oe). 

One can even identify the field induced turnaround temperature $T_2$$^*$, assosiated with the second transformation $T_{f2}$ in Fig. 6. The high field ($H$ \textgreater\ 2 kOe) $M_{FCC}$($T$) curves in Nd$_{0.75}$Gd$_{0.25}$Rh$_{3}$B$_2$ (see Fig. 4 in Ref. [11]) continue to imbibe the presence of two transitions (at $T_{f1}$ and $T_{f2}$). Fig. 7 shows the $M_{FCC}$($T$) curves in $H$= 30 kOe, 50 kOe and 70 kOe in Nd$_{0.8}$Tb$_{0.2}$Al$_2$. $T_c$ and $T_{f2}$ values have been marked in Fig. 7. Note that the $M_{FCC}$($T$) curves in $H$ $\apprge$ 50 kOe could be considered as oblivious to the transformation setting in at $T_{f2}$ in lower fields. This could also be termed as consistent with the absence of any fingerprint of the lower temperature transition in $C_p$($T$) curve in $H$ = 50 kOe in Fig. 5. 

\begin{figure}[h]
\includegraphics[width=0.45\textwidth]{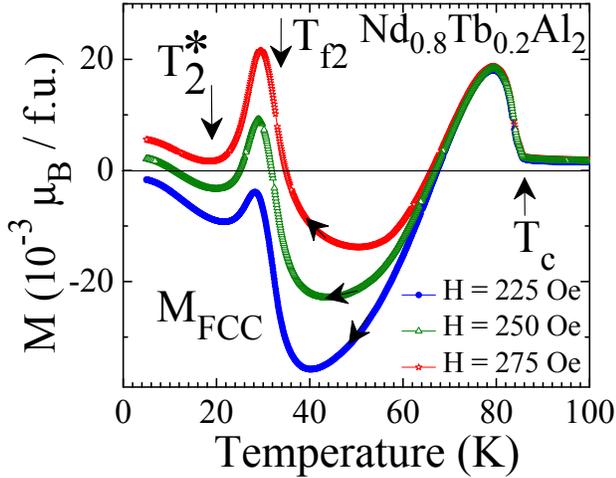}
\caption{\label{fig4}(Color online) Field cooled cool down (FCC) magnetization curves in Nd$_{0.8}$Tb$_{0.2}$Al$_2$ in $H$ = 225 Oe, 250 Oe and 275 Oe. Note the multiple crossovers of the $M$ = 0 axis by $M$$_{FCC}$($T$) curve in $H$ = 250 Oe.}
\end{figure}

\begin{figure}[h]
\includegraphics[width=0.45\textwidth]{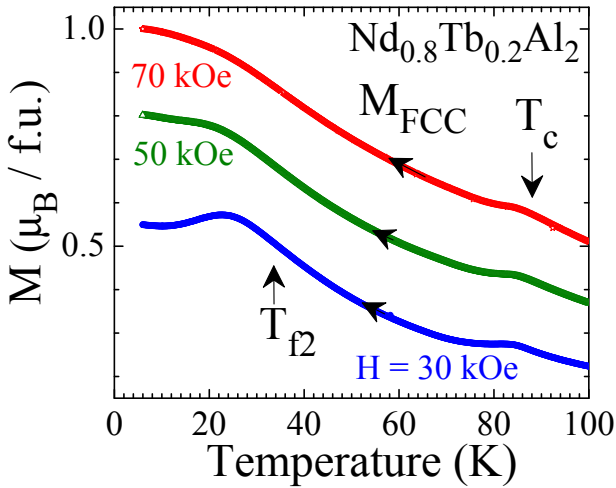}
\caption{\label{fig4}(Color online) $M_{FCC}$($T$) curves in $H$ = 30 kOe, 50 kOe and 70 kOe in Nd$_{0.8}$Tb$_{0.2}$Al$_2$. The locations of $T_c$ and $T_{f2}$ determined from the low field data are identified.}
\end{figure}

\section{Discussion}

\begin{figure}[h]
\includegraphics[width=0.45\textwidth]{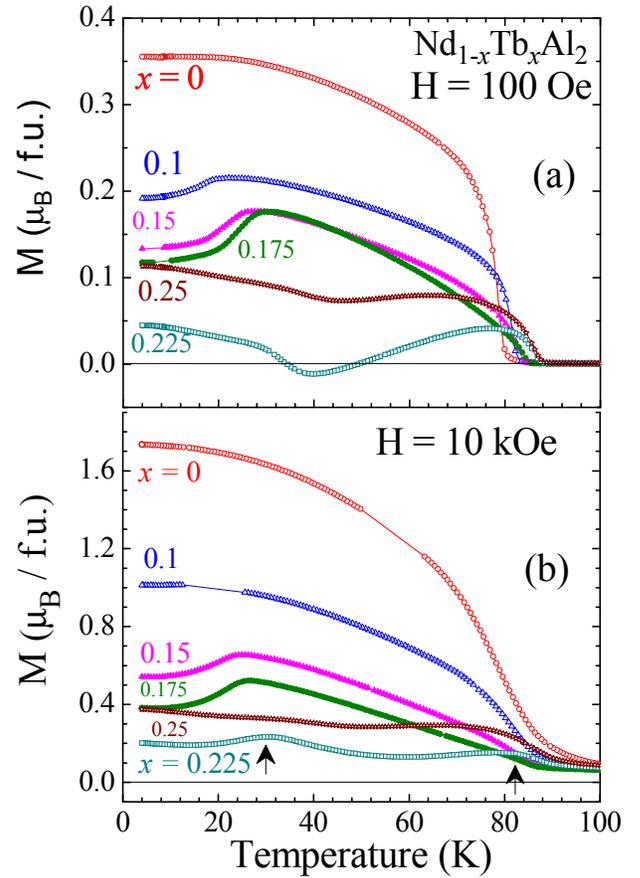}
\caption{\label{fig4}(Color online) FCC magnetization curves measured in (a) 100 Oe and (b) 10 kOe in Nd$_{1-x}$Tb$_{x}$Al$_2$ series. The progressive evolution in behavior and the emergence of second transformation below 40 K can be noted in both the panels. In Fig. 8(a), multiple crossings of the $M$ = 0 axis are evident for $x$ = 0.225. In Fig. 8(b), the two peaks (identified by arrows for $x$ = 0.225) centred around 82 K and 30 K can also be clearly noted.}
\end{figure}

The magnetic compensation behavior shown in Fig. 2 (a) in Nd$_{0.8}$Gd$_{0.2}$Al$_2$ arises from the pseudo-ferrimagnetic character of the admixed intermetallic and the differences in the temperature dependences of the Nd$^{3+}$ ions beloging to the first half of the 4$f$-RE series and the $S$-state (i.e., $L$ = 0) Gd$^{3+}$ ions easily account for it. The discovery of the repeated magnetic compensation in Nd$_{0.8}$Tb$_{0.2}$Al$_2$ has been an unexpected finding, our belief in it as an authentic behavior is firmed up by the broad similarities in the magnetization, ac-susceptibility, dc electrical resistance and specific heat data in this system with those in the admixed intermetallic  Nd$_{0.75}$Gd$_{0.25}$Rh$_{3}$B$_2$. The basis of repeated crossover of $M$ = 0 axis in dc magnetization data is the two magnetic transitions involving antiferromagnetically coupled magnetic moments of Nd$^{3+}$ and Gd$^{3+}$/Tb$^{3+}$ in Nd$_{1-x}$Gd$_{x}$Rh$_{3}$B$_2$ /  Nd$_{1-x}$Tb$_{x}$Al$_2$ alloys. To further support these assertions, we first draw attention to the low ($H$ = 100 Oe) and high ($H$ = 10 kOe) $M_{FCC}$($T$) curves for the series of Nd$_{1-x}$Tb$_{x}$Al$_2$ alloys (0 $\leq$ x $\leq$ 0.25) in Figs. 8(a) and 8(b). The progressive replacements of Nd$^{3+}$ ions by Tb$^{3+}$ ions lowers the magnetization signal in the ordered state, and between $x$ = 0.175 and $x$ = 0.225, the emergence of two distinct transitions (near 85 K and near 40 K) can be clearly noted in the sets of curves in both the panels of Fig. 8. In lower field of 100 Oe, the multiple crossovers of $M$ = 0 axis can be noted in $x$ = 0.225 sample (campare this with the corresponding response in $x$ = 0.20 sample in Fig. 2(a)). In the higher field of 10 kOe, $M_{FCC}$($T$) response in $x$ = 0.20 alloy (see Fig. 2(b)) lies in between the corresponding responses for $x$ = 0.175 and $x$ = 0.225 alloy in Fig. 8(b). At $x$ = 0.25, the balance between the contributions from Nd$^{3+}$ ions and Tb$^{3+}$ ions shifts towards the latter, and the interesting details of the repeated magnetic compensation behavior become difficult to discern. 

Some of us \cite{Kulkarni1} have recently succeeded in growing a large single crystal (4 cm in length and 3 mm in diameter) near the stoichiometry Nd$_{0.8}$Tb$_{0.2}$Al$_2$ like the crystal of Nd$_{0.75}$Ho$_{0.25}$Al$_2$\cite{Kulkarni2}. Figure 9 shows the $M_{FCC}$($T$) curves in $H$ = 125 Oe and 150 Oe for $H$ $\parallel$ [100] in a tiny crystal cut from it. The repeated magnetic compensation behavior is vivid in this panel. The values of $T_c$, $T_{f2}$, $T_{comp1}$ and $T_2$* have been identified in Fig. 9. 

\begin{figure}[h]
\includegraphics[width=0.45\textwidth]{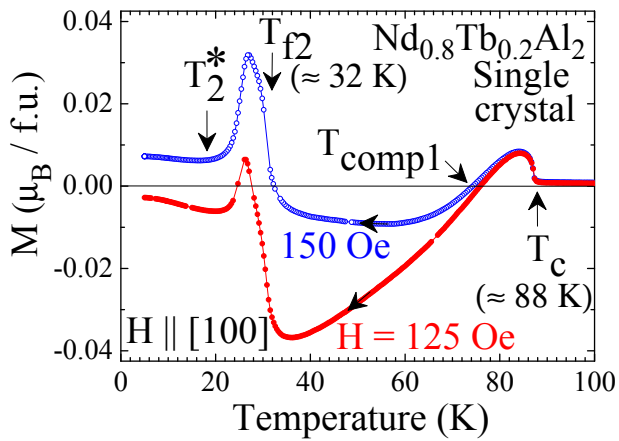}
\caption{\label{fig7}(Color online) $M_{FCC}$($T$) curves in $H$ = 125 Oe and 150 Oe in a single crystal of Nd$_{0.8}$Tb$_{0.2}$Al$_2$ ($H$ $\parallel$ [100]). The magnetic ordering temperature ($T_c$) and the onset temperature of second transformation (at $T_{f2}$) are marked. $T_{comp1}$ and $T_2$* values are also identified from the two $M_{FCC}$($T$) curves.}
\end{figure}

We are tempted to speculate on the cause of the repeated magnetic compensation behavior at some specific compositions in the admixed Nd$_{1-x}$Tb$_{x}$Al$_2$ and Nd$_{1-x}$Gd$_{x}$Rh$_{3}$B$_2$ intermetallics. Gross inhomogenity at the macroscopic level in the samples is ruled out from the cleanliness (i.e., very narrow widths, with no sign of any impurity peaks) of the x-ray diffraction patterns of the finely powdered specimen (see, e.g., inset of Fig. 1). The single crystal data in Fig. 9 for Nd$_{0.8}$Tb$_{0.2}$Al$_2$ also supports the genuineness of the observed phenomenon. However, statistical fluctuations at the atomic level, which maintain (1-$x$):$x$ ratio of Nd$^{3+}$ and Tb$^{3+}$/Gd$^{3+}$ ions at a larger length scale, cannot be fully ruled out as a cause of bimodal distribution in the time scales of magnetic dynamics centred around clusters of Nd$^{3+}$ and Gd$^{3+}$/Tb$^{3+}$ ions in the respective systems. The intertwinned chains of clusters of two types, the dynamics of one of which slows down near the upper transition (at $T_{c}$), followed by complete freezing of the dynamics of all the magnetic ions in the entire sample below the onset of the lower transition (at $T_{f2}$), could account for the repeated magnetic compensation behavior in Nd$_{0.75}$Gd$_{0.25}$Rh$_{3}$B$_2$ and Nd$_{0.8}$Tb$_{0.2}$Al$_2$ alloys. The details of the slow down of magnetic dynamics in the two systems are likely to be different, considering that the freezing in of the spin-disorder resistivity is conspicuous at $T_c$ in Nd$_{0.8}$Tb$_{0.2}$Al$_2$, as compared to its conspicuousness at $T_{f2}$ in Nd$_{0.75}$Gd$_{0.25}$Rh$_{3}$B$_2$. The important differences in the specific heat responses in the two systems also call for a microscopic explanation. 

It is hoped that the new results presented here would spur a simulation study of the admixed rare earth intermetallic systems, where long range RKKY interaction (via the conduction electrons) mediates the coupling between the 4$f$-spins of dissimilar rare earth ions. The x-ray magnetic circular dichroism (XMCD) studies in admixed Nd$_{1-x}$Tb$_{x}$Al$_2$ and Nd$_{1-x}$Gd$_{x}$Rh$_{3}$B$_2$ series could also be useful in tracking the temperature dependences of 4$f$-spins and 4$f$-orbital contributions of individual rare earth ions in the entire sample. Any gross difference in the inhomogeniety/dynamics would show up in the thermal evolution of the sub-parts of the magnetic moments of the 4$f$-rare earth ions.  

\section{Summary and Conclusion}

To summarize, we have presented heighlights of the results of our investigations in admixed rare earth alloys with near-zero net-magnetization in the series Nd$_{1-x}$Gd$_{x}$Al$_2$ and Nd$_{1-x}$Tb$_{x}$Al$_2$. The interest in the stoichiometries imbibing near-zero magnetization characteristic had been rekindled in contemporary times, as such samples concurrently possess large spin polarization in the conduction band, which can find niche application potential in spintronics\cite{Adachi1}. The discovery of repeated magnetic compensation in Nd$_{0.75}$Gd$_{0.25}$Rh$_3$B$_2$ had been an unexpected finding\cite{Kulkarni}. This motivated us to search for such a behavior in zero magnetization stoichiometries in different ferromagnetic rare earth series. Our present successful identification of repeated magnetic compensation in Nd$_{0.8}$Tb$_{0.2}$Al$_2$, alongwith the earlier results in Nd$_{0.75}$Gd$_{0.25}$Rh$_3$B$_2$\cite{Kulkarni} emphasize that the repeated/multiple magnetic compensation phenomenon should be considered as a generic behavior for the admixture of the two rare earth ions belonging to different halves of 4$f$-series at an appropriate stoichiometry. However, the details of manifestation of such phenomenon could be different in different series, depending on the crystal structure and the magnetic anisotropy of the rare earth ions involved in the composition. A clear understanding of repeated magnetic compensation phenomenon would need comprehensive theoretical modeling of the admixed systems. 

Acknowledgements

We would like to thank R. Kulkarni and N. Kulkarni for their help in some of the measurements.


\begin{thebibliography}{99}

\bibitem{Wallace}W. E. Wallace, E. Segal, Rare Earth Intermetallics (Academic Press, New York, 1973). 
\bibitem{Purwins}H. G. Purwins and A. Leson, Advances in Physics, 39, 309 (1990) and references therein. 
\bibitem{Williams}H. J. Williams, J. H. Wernick, E. A. Nesbitt, R. C. Sherwood, J. Phys. Soc. of Jpn. 17-B1, 91 (1962). 
\bibitem{Swift}W. M. Swift, W. E. Wallace, J. Phys. Chem. Solids 29, 2053 (1968).
\bibitem{Grover1}A. K. Grover, S. K. Malik, R. Vijayaraghvan, K. Shimizu, J. Appl. Phys. 50, 7501 (1979).
\bibitem{Grover2}A. K. Grover, D. Rambabu, S. K. Dhar, S. K. Malik, R. Vijayaraghavan, G. Hilscher, H. Kirchmayr, Proceedings of Annual (1983) Department of Atomic Energy (DAE) Nuclear Physics and Solid State Physics Symposium (India) 26C, 252 (1983).
\bibitem{Adachi1}H. Adachi and H. Ino, Nature 401, 148 (1999). 
\bibitem{Adachi2}H. Adachi, H. Ino, H. Miwa, Phys. Rev. B 59, 11445 (1999). 
\bibitem{Adachi3}H. Adachi, H. Ino, H. Miwa, Phys. Rev. B 56, 349 (1997).
\bibitem{Adachi4}H. Adachi, H. Kawata, H. Hashimoto, Y. Sato, I. Matsumoto, Y. Tanaka, Phys. Rev. Lett. 87, 127202 (2001).
\bibitem{Kulkarni}P. D. Kulkarni, U. V. Vaidya, V. C. Rakhecha, A. Thamizhavel, S. K. Dhar, A. K. Nigam, S. Ramakrishnan and A. K. Grover, Phys. Rev. B 68, 064426 (2008). 
\bibitem{Obiraki}Y. Obiraki, H. Nakashima, A. Galatanu, T. D. Matsuda, Y. Haga, T. Takeuchi, K. Sugiyama, K. Kindo, M. Hagiwara, R. Settai, H. Harima and Y. Onuki, J. Phys. Soc. Jpn. 75, 064702 (2006).
\bibitem{Ramirez}A. P. Ramirez,  L. F. Schneemeyer and  J. V. Waszczak, Phys. Rev. B 36, 7145 (1987). 
\bibitem{Kulkarni1} P. D. Kulkarni, A. Thamizhavel, Unpublished. 
\bibitem{Kulkarni2} P. D. Kulkarni, A. Thamizhavel, V. C. Rakhecha, A. K. Nigam, P. L. Paulose, S. Ramakrishnan and A. K. Grover, Europhys. Lett. 86, 47003 (2009). 
\end{thebibliography}
\end{document}